\documentclass{aa}
\usepackage{graphics}
\begin{document}

   \thesaurus{06     
              (08.02.1;  
               02.02.1;  
               13.25.5;  
               08.09.2;  
               08.14.2)} 

\title{Discovery of metal line emission from the Red star 
in IP Peg during outburst maximum}

\titlerunning{Red star emission in IP Peg}
\authorrunning{Harlaftis}

\vspace{1cm}

   \author{E. Harlaftis\inst{1} \thanks{\emph{Visiting Scientist:} School of 
           Physics and Astronomy, University of St. Andrews, KY16  9SS, UK}
                     }

   \offprints{E. Harlaftis}

   \institute{Astronomical
           Institute, National Observatory of Athens, Lofos Nymfon, P.O. Box
           20048, Athens 11810, Greece\\email: ehh@astro.noa.gr)
	}

   \date{Received, 6 April 1999; accepted, 12 May 1999}

   \maketitle

   \begin{abstract}

Observations of the eclipsing dwarf nova IP Peg during outburst reveal
metal lines in emission, such as Mg{\small~II} 4481 \AA. \ Analysis using
Doppler tomography locates emission of helium and metal lines on the inner
Roche lobe of the secondary star. Such multi-line Roche-lobe imaging
presents a new tool in mapping the red star's ionization structure.

\keywords{cataclysmic variables, accretion disc, IP Pegasi}

   \end{abstract}

%

\section{Introduction}

Cataclysmic variables are interacting binaries where a white dwarf and
a red dwarf orbit  each other within a few  hours.  Line emission from
the red star is now regularly detected  (Harlaftis and Marsh 1996, and
references therein).  During    outburst of the  eclipsing dwarf  nova
IP~Peg, irradiation  from the hot central  regions of the disc is most
likely  responsible for  the line emission   located  on the red  star
(Marsh and Horne   1990).  During quiescence, H$\alpha$  line emission
from the red star of IP Peg is transient  and its origin is unresolved
(Harlaftis  et al.   1994).   The fast  rotation of  the  red stars in
cataclysmic variables and the  regular irradiation of their atmosphere
by   the hot accretion  disc present   a  physical situation which may
affect, in the  long   term, the  atmospheric  stratification of   the
companion star and   its subsequent evolution. Techniques  for mapping
either the surface  of cool  single  stars from the  absorption  lines
(Cameron 1999) or the surface of the red star in cataclysmic variables
from  the emission  lines (Marsh and  Horne  1988; Rutten and  Dhillon
1994) have been  developed.  These techniques can  be  used in probing
the ionization structure  of the  upper atmosphere  of the red   star.
Here, we report on spectrophotometric observations of IP Peg, obtained
with the 2.5m INT  at La Palma,   during maximum of the  November 1996
outburst, which  were aimed to   probe the structure  of the, recently
discovered, spiral arms in   the  disc of  IP  Peg (Harlaftis  et  al.
1999).  As  a by-product of the  observations, we discover metal lines
in emission from the secondary star.

\begin{figure}
\resizebox{\hsize}{!}{\rotatebox{-90}{\includegraphics{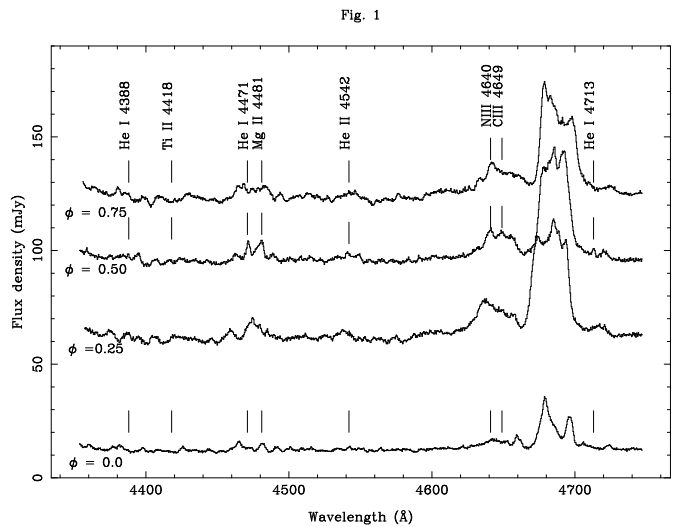}}}
\caption[]{Average spectra of IP Peg in outburst 
at characteristic binary phases (as marked).
The emission lines are sharper at phase 0.5, when the inner side of the red 
dwarf passes through the line-of-sight. During eclipse, 
structure on the continuum is significantly suppressed and the
emission-line flux is still detectable.
} 
\end{figure}

\section{Doppler tomography}

\begin{figure*}
\resizebox{\hsize}{!}{\rotatebox{-90}{\includegraphics{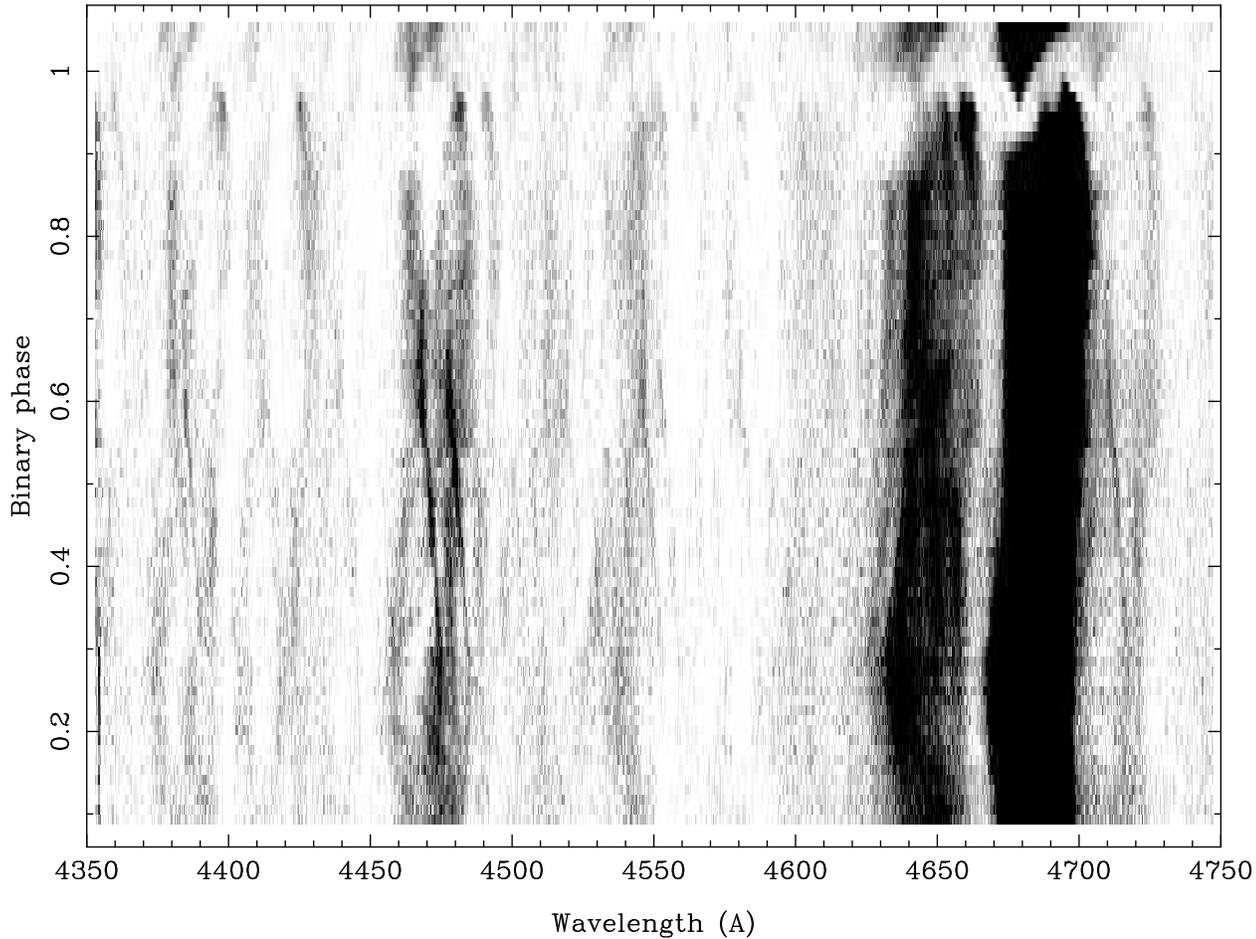}}}
\caption[]{Trailed spectra over the 4350--7450\AA \ range scaled, so that
the  motion of  weak lines  is revealed (0--10   mJy).  All  the lines
identified in Fig. 1  can  be traced here,  such as  the Ti{\small~II}
line  at 4418\AA. The  trailed  spectra of  the  Bowen blend are  very
similar to the He{\small~II} line.  The velocity  relative to the line
centre along the horizontal axis is plotted and the binary phase along
the   vertical    axis.  The intensity   scale      is 0-10 mJy (greyscale bar)
.}
\end{figure*}

IP Peg   is a well-studied,  double-eclipsing  dwarf nova, which shows
semi-periodic outbursts every $\sim$3 months.     For details on   the
observations and data reduction see Harlaftis et  al. (1999).  Average
spectra, in the range 4354--4747 \AA, \  at four characteristic binary
phases are displayed  in Fig.  1.  In  addition to the Bowen blend and
He{\small~II} 4686 lines, weak He{\small~I} lines at 4388, 4471, 4713
\AA \, Mg{\small~II} 4481 \AA \ and Ti{\small~II} 4418 \AA \ are 
also  observed (see also trailed  spectra  in Fig.   2).  These  lines
display  a sharp peak  at phase   0.5  indicating a component from  the
companion star.  We adopt the binary ephemeris from Wolf et al. (1993)
$ T_{o}(HJD)  = 2445615.4156(4) +  0.15820616(4)~E$, where  $T_{o}$ is
the inferior conjunction of the white dwarf. 

The trailed spectra over the full wavelength range are shown in Fig. 2
with the aim to  display the motion of the  weak lines  (the intensity
scale is adjusted so that He{\small~II}  line appears saturated).  The
disc   and red star    emission components are   seen  in the lines of
He{\small~I}  4388, He{\small~I} 4472,  Mg{\small~II}  4481, the Bowen
blend and the  He{\small~I} 4713. The red star  component is the sharp
`S'-wave moving from red to blue at phase 0.5.   It can also be traced
in the He{\small~II} 4542 and the Ti{\small~II} 4418 lines.  Note that
the Mg{\small~II} `S'-wave component  disappears earlier (binary phase
0.7) than   that of the neighbouring   He{\small~I} 4472 (binary phase
0.75).

We  reconstruct the  Doppler images of   the emission lines using  the
trailed  spectra  (Marsh and   Horne  1988).  A  Doppler  image is the
reconstruction of the emission line distribution in velocity space and
has been particularly successful in resolving the location of emission
components such as  the red star (IP Peg;  Harlaftis et al. 1994), the
gas stream (OY Car in outburst;  Harlaftis and Marsh 1996), the bright
spot (GS2000+25; Harlaftis et al.  1996) and spiral waves in the outer
accretion disc (Steeghs,  Harlaftis and Horne  1997; Harlaftis et  al.
1999).  We built  the Doppler images of the  emission lines (see above
references for the procedure)  and, after subtracting the axisymmetric
disc emission, we can zoom  onto the Roche  lobe of the red star (Fig.
3  from  left   to right,  high-ionization    to low-ionization lines,
He{\small~II} 4686, He{\small~I} 4388, He{\small~I} 4472, He{\small~I}
4713, Mg{\small~II} 4481).

\begin{figure*}
\resizebox{\hsize}{!}{\rotatebox{-90}{\includegraphics{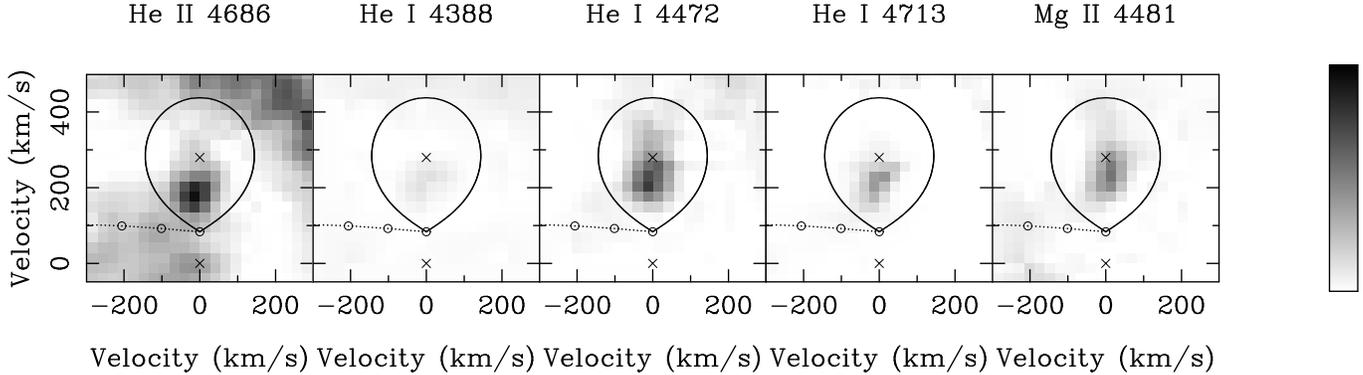}}}
\caption[]{The irradiated Roche lobe of the red dwarf in IP Peg 
is imaged at various wavelengths. From left to right, the ionization
potential decreases. These are the Doppler images
zoomed on the red dwarf. The irradiated area may be moving to the polar 
regions with decreasing ionization potential.}
\end{figure*}

\section{Line emission from the Red star}

The line emission, on the Doppler images of Fig. 3, is stronger on the
side of the red  star facing the  white  dwarf and further  it weakens
towards the equator  of the red star,  indicative of screening by  the
disc.  The He{\small~I} maps show a relative  strength similar to that
seen  in diffuse nebulae  (weaker in 4388\AA,  \ stronger in 4472 \AA;
Kaler 1976). We measured      the  velocity locations of    the   peak
intensities  in the Doppler  images using Gaussian fitting.  There may
be a systematic shift towards the L$_{1}$ point with higher-ionization
potential (see  Table 1). As a consistency test for the properties of the
``spots'' being realistic, the velocity widths of the
irradiated sites are indeed less than the rotational broadening of
the companion star, $\upsilon
\sin  i = 146$   km s$^{-1}$ (1$\sigma$),  which  is obtained from  the
relation

\[ \frac{\upsilon \sin i}{K_{\rm c}} =
0.462 \left[ (1+q)^{2} ~q \right] ^{1/3} .  \]

where $q=0.58\pm0.10$, and $K_{\rm c}  = 280\pm2$ km s$^{-1}$ (Wolf et
al. 1998). The measured 
relative shifts between the wavelength-dependent irradiated sites
are with respect to zero velocity (binary's centre), therefore 
any uncertainties in the system parameters 
do not alter our conclusion.

In the past,  similar emission has  been interpreted as irradiation of
the inner  side of the red  star by soft X-ray  photons emitted by the
boundary layer (Harlaftis \& Marsh  1996 and references therein).  The
Roche  lobe maps may suggest  that there is temperature foreshortening
or that  the shadow cast  by the disc  on the companion star decreases
with  higher    energy   photons    (Mg{\small~II},   He{\small~   I},
He{\small~II}).   Indeed,  the  disc  thickness  may  hinder efficient
irradiation around L$_{1}$ relative to the polar regions.

\begin{table}
\caption{Red star emission}
\begin{tabular}{lccc}
ion          & $\lambda$ &eV& km s$^{-1}$\\
He{\small~II}& 4686 &54.4&+185$\pm$53 \\
He{\small~I} & 4388 &24.6&+216$\pm$32 \\
He{\small~I} & 4472 &24.6&+213$\pm$39 \\
He{\small~I} & 4713 &24.6&+218$\pm$44 \\
Mg{\small~II}& 4481 &15.0&+240$\pm$54 \\
\end{tabular}
\end{table}

In this way,  the red star emission can  also  be used to measure  the
thickness of the disc as seen by the soft X-rays which excite the line
emission, independently  of X-ray data.  From the  values in  Table 1,
the L$_{1}$ region  may mainly  be clear  of   emission around 50   km
s$^{-1}$ (or 2 pixels) from the L$_{1}$.   This corresponds to a Roche
lobe height of $\sim~30$  \%, or $\sim~0.10$ $\alpha$, where  $\alpha$
is the binary separation (assuming q=0.6 and use of Table 3-1 in Kopal
1959).  Therefore, there is potential to  probe the vertical structure
of the disc  with higher quality  data.  Moreover, disc  contamination
and  small-scale blurring  caused  by  the Doppler  tomography  (which
assumes isotropic emission from the orbital  plane) are significant at
that level and only  improvement  of the Doppler  tomography technique
combined with   higher-resolution   data will   clarify   better   the
ionization zones on the Roche lobe and the vertical disc height.

\begin{acknowledgements}

The   data reduction  and  analysis   was partly   carried out at  the
St. Andrews STARLINK node.  Use of software developed  by T.  Marsh is
acknowledged.       ETH was   supported    by the     TMR contract  RT
ERBFMBICT960971 of the European Union.  ETH was partially supported by
a joint    research programme between the    University of Athens, the
British Council at Athens and the University of St. Andrews.

\end{acknowledgements}

\label{lastpage}

\end{document}